\documentclass{INTERSPEECH2023}

\interspeechcameraready

\title{PIAVE: A Pose-Invariant Audio-Visual Speaker Extraction Network}
\name{Qinghua Liu$^1$$^,$$^2$, Meng Ge$^2$$^,$$^3$$^\ast$\thanks{$^\ast$Corresponding author}, Zhizheng Wu$^2$, Haizhou Li$^1$$^,$$^2$$^,$$^3$}
\address{
  $^1$Shenzhen Research Institute of Big Data, Shenzhen, China\\
  $^2$School of Data Science, The Chinese University of Hong Kong, Shenzhen, China\\
  $^3$ Department of Electrical and Computer Engineering, National University of Singapore, Singapore}
\email{\{liuqinghua,gemeng,wuzhizheng,haizhouli\}@cuhk.edu.cn}

\begin{document}

\maketitle
 
\begin{abstract}
It is common in everyday spoken communication that we look at the turning head of a talker to listen to his/her voice. Humans see the talker to listen better, so do machines. However, previous studies on audio-visual speaker extraction have not effectively handled the varying talking face. This paper studies how to take full advantage of the varying talking face. We propose a Pose-Invariant Audio-Visual Speaker Extraction Network (PIAVE) that incorporates an additional pose-invariant view to improve audio-visual speaker extraction.
Specifically, we generate the pose-invariant view from each original pose orientation, which enables the model to receive a consistent frontal view of the talker regardless of his/her head pose, therefore, forming a multi-view visual input for the speaker.
Experiments on the multi-view MEAD and in-the-wild LRS3 dataset demonstrate that PIAVE outperforms the state-of-the-art and is more robust to pose variations.

\end{abstract}

\section{Introduction}
The human brain has a remarkable ability to focus auditory attention on a particular voice by masking out the acoustic background in the presence of multiple speakers and background noises~\cite{cherry1953some}, that is called cocktail party effect~\cite{bronkhorst2000cocktail}. 
With the advent of deep learning~\cite{wang2018supervised}, neural approaches become increasingly popular in solving the cocktail party problem. 
The speaker extraction is one of such solutions that extracts the target speaker of interest from a multi-talk environment. It relies on an auxiliary cue to direct the attention towards the target speaker. 
The auxiliary cue may take different forms, such as  pre-recorded reference speech~\cite{wang2019voicefilter,xu2020spex,ge2020spex+}, speech-synchronized video clips~\cite{wu2019time,gao2021visualvoice,liu2023limuse,li22ba_interspeech}, and the target speaker's spatial location~\cite{gu2019neural,ge2022spex}. Notably, visual information is highlighted as a particularly useful cue for its immunity to acoustic noise and competing speakers~\cite{michelsanti2021overview} and is more informative than an audio cue~\cite{pan2022usev}.

\begin{figure}[t]
	\centering
	\includegraphics[width=\linewidth]{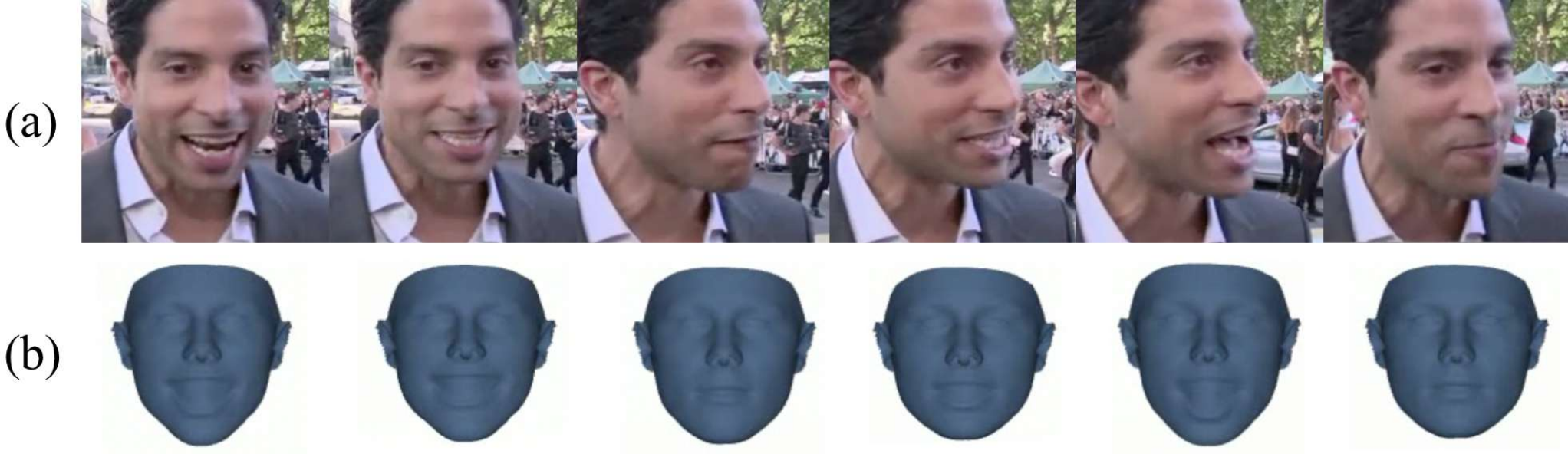}
	\caption{(a) An illustration of the pose variation in daily conversations. 
                (b) Pose-invariant faces corresponding to the above row of faces.}
	\label{fig:pose}
\end{figure}

Although recent audio-visual speaker extraction (AVSE) systems have demonstrated significant improvements in performance across a variety of standard datasets, the impact of head pose variations has not been studied. Head pose variations~\cite{baldassare1978human} as shown in Figure~\ref{fig:pose}-(a) will result in changes in the orientation of the speaker's face in the camera view.
Such variations adversely impact visually-assisted speech processing systems such as lipreading~\cite{cheng2020towards} and audio-visual speaker extraction.

\begin{figure*}[htb]
	\centering
	\includegraphics[width=0.83\linewidth]{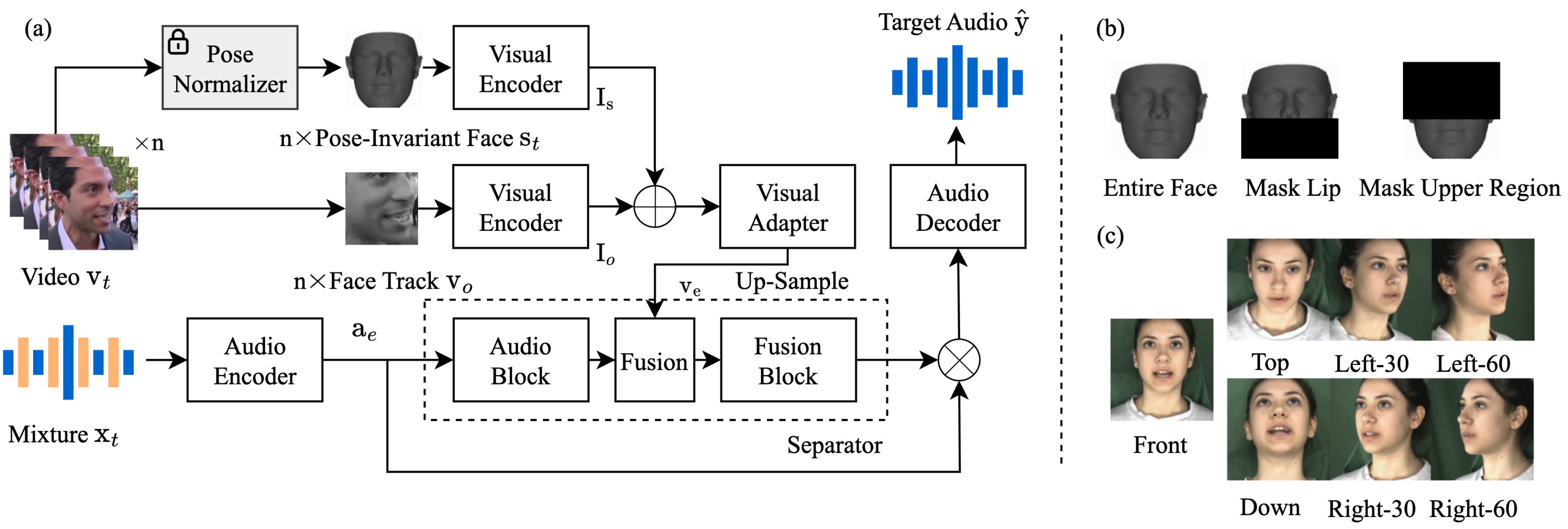}
	\caption{(a) The overall architecture of the proposed PIAVE network. (b) An illustration of masking regions in Table~\ref{tb:multi_view} and \ref{tb:ablation}. (c) Multi-view illustration of the MEAD dataset.}
	\label{fig:overview}
\end{figure*}


We argue that an invariant pose could help audio-visual speaker extraction. Hence,
we propose a Pose-Invariant Audio-Visual Speaker Extraction Network, namely PIAVE.
This approach involves generating pose-invariant view (front view, in this paper) faces from the original face track, enabling that the model receives a consistent frontal view of the talker regardless of head pose variations.
Furthermore, the model benefits from multi-view observation of the talking face, namely the generated pose-invariant view and the original pose orientation, thereby aiding in the more accurate identification of the target speaker and obtaining better performance in speaker extraction.

The contributions of this paper can be summarized as follows.
Firstly, PIAVE represents the first step towards addressing the pose variation problem in AVSE.
Secondly, we effectively generate the corresponding pose-invariant face for any given face image input, as shown in Figure~\ref{fig:pose}-(b), which ensures stable visual input for the model.
Lastly, when only one camera is used, PIAVE benefits from the multi-view observation of the target speaker, outperforming the state-of-the-art.

\section{Speaker Extraction with Visual Cues}
Figure~\ref{fig:overview} is an illustration of the workflow of the proposed PIAVE network during run-time inference.
The network takes an audio mixture $\mathrm{x}_t$ and a video sequence of the target speaker’s faces $\mathrm{v}_t$ as inputs. It could be described as:
\begin{equation}
\mathrm{\hat{y}} = \text{Decoder}^a(\textrm{Separator}(\mathrm{a}_e , \mathrm{v}_e) \odot \mathrm{a_e}),
\end{equation}
where $\mathrm{\hat{y}}$ is the predicted target audio, $\odot$ is an operation for element-wise multiplication, $\mathrm{a}_e$ and $\mathrm{v}_e$ is encoded representations of the audio mixture and video sequence. 
The network consists of four parts: encoder, separator, decoder and pose normalizer. The pose normalizer will be described in Section~\ref{section:ff}.

The audio encoder converts the audio mixture into a spectrum-like representation in the time domain, and the audio decoder is used to transform the masked encoder representation to the target audio following~\cite{luo2019conv}.

The visual encoder extracts visual embeddings, which model the temporal synchronization and interaction between visemes and speech.
It has a 3D convolutional layer followed by an 18-layer ResNet~\cite{he2016deep}, which is pretrained on the lipreading task following~\cite{afouras2018deep}.
It takes a sequence of pose-invariant faces and the original face track as input and outputs the fixed dimensional feature vectors $\mathrm{I}_s\in\mathbb{R}^{d \times n}$ and $\mathrm{I}_o\in\mathbb{R}^{d \times n}$ for the two video streams, where $d$ denotes the feature dimension of the visual embedding and $n$ denotes the number of frames in each video stream.
These feature vectors are fused together using addition and passed through a visual adapter that models the temporal dependencies across each frame to capture the temporal dynamics of the visual input.
Moreover, since the time resolution of the video stream and the audio stream is different, we upsample the visual embeddings along the temporal dimension to synchronize the audio stream and the video stream by nearest neighbor interpolation.

The separator is designed for estimating a mask to let pass the target speaker, and filter out others, conditioned on the encoded audio and visual features.
To capture the long-term dependencies in the encoded audio features, the audio block employs a TCN network consisting of multiple TCN blocks, each with a depth-wise separable convolution, PReLU activation, and layer normalization operation.
The output of the audio block is gathered with the encoded visual features $\mathrm{v}_e$ through concatenation and then we use two repeats of the TCN network to process the fused audio-visual feature.
A convolutional layer is followed by the fusion block to generate the mask of the target speaker. It is then multiplied by the encoded audio representations of the mixture $\mathrm{a}_e$ to obtain the target audio $\mathrm{\hat{y}}$.

During training, the Scale-Invariant Signal-to-Distortion ratio (SI-SDR)~\cite{le2019sdr} between the predicted target audio $\hat{y}$ and the ground-truth target audio $\mathrm{y}$ is used to optimize the network from end to end:
\begin{gather}
    \label{eq:sisdr}
    \mathcal{L}_{\text{SI-SDR}} = -\rho(\hat{y}, y), \\
    \rho(\hat{y}, y) = 20 \log_{10}\frac{||(\hat{y}^T y / y^T y) \cdot y||}{||(\hat{y}^T y / y^T y) \cdot y - \hat{y}||}.
\end{gather}


\section{Head Pose Normalization}
\label{section:ff}
\subsection{Problem formulation}
The variation of head poses has always been a challenge in both visual-only tasks and multi-modal processing.
With large head pose variations, the intra-person variance of head representation may drastically increase, sometimes even exceeding inter-person variance.
Recent research has shown that pose normalization consistently boosts the performance of face recognition~\cite{zhu2015high,he2019deformable}. It should be noted that expression-free pose normalization is required for face recognition tasks.
However, in AVSE, lip movements and facial expressions that contribute to speech production correspond with phonetic content and have a strong impact on the ability of humans to focus their auditory attention~\cite{mcgurk1976hearing, schultz2017biosignal}.
For this reason, expression-preserving pose normalization is required in AVSE.

In this paper, we propose pretraining the pose normalization module (i.e. pose normalizer in Figure~\ref{fig:overview}-(a)) on the 3D face alignment and reconstruction task and keeping it frozen during the speaker extraction model training.
The pose normalizer is expected to generate the expression-preserving pose-invariant face for any given face image input.
The challenge is that large pose diversity makes it hard to distinguish facial expression variations and increases the modeling difficulty.
To address this problem, we draw inspiration from~\cite{ruan2021sadrnet} and propose to disentangle the head pose and facial expression to reduce the complexity of the problem and make it more tractable.
In the following sections, we will introduce the pretraining setup for this dual face regression network and the approach we have adopted for head pose normalization.

\subsection{Facial geometry representation}
We separate the 3D face geometry into pose, mean shape, and deformation as:
\begin{equation}
\label{eq:transformation}
\mathrm{G}=f\times \mathrm{R} \times \mathrm{S} + \mathrm{t},
\end{equation}
\begin{equation}
\mathrm{S}=(\mathrm{\overline{S}}+\mathrm{D}).
\end{equation}
where $\mathrm{G}\in \mathbb{R}^{3 \times m}$ is the 3D mesh of a specific face with $m$ vertices. The pose parameters consist of the scale factor $f$, the 3D rotation matrix $\mathrm{R}\in \mathbb{R}^{3 \times 3}$ and the 3D translation $\mathrm{t}\in \mathbb{R}^{3}$.
$\mathrm{S}\in \mathbb{R}^{3 \times m}$ represents the pose-invariant face shape, which is disentangled into the mean shape template $\mathrm{\overline{S}}\in \mathbb{R}^{3 \times m}$ from~\cite{paysan20093d} and the deformation $\mathrm{D}\in \mathbb{R}^{3 \times m}$ between the actual shape $\mathrm{S}$ and the mean shape $\mathrm{\overline{S}}$.

\subsection{Dual face regression network}
Following the approach proposed in~\cite{ruan2021sadrnet}, we adopt a joint regression strategy to simultaneously estimate the pose-invariant face $\mathrm{S}$ and the pose-dependent face $\mathrm{P}$. Then, we employ a self-alignment module denoted by $\phi$ to estimate face pose from the estimated faces:
\begin{equation}
\phi(\mathrm{P}, \mathrm{S})=f, \mathrm{R}, \mathrm{t}.
\end{equation}
Based on the estimated face pose $\phi(\mathrm{P}, \mathrm{S})$ and pose-invariant face $\mathrm{S}$, we could reconstruct the final shape $\mathrm{G}$ via transformation defined in Eq.~(\ref{eq:transformation}).

The joint regression of the pose-invariant face $\mathrm{S}$ and pose-dependent face $\mathrm{P}$ is complimentary.
For one thing, it helps to separate the effects of non-rigid facial changes due to expression from those resulting from rigid facial changes due to head pose.
For another, the learning of pose-dependent face is easy to over-fit to pose and under-fit to facial expressions as the pose variations bring much greater point-to-point distances than the expression variations, resulting in poor reconstruction of expression in various views.
By introducing the pose-invariant face $\mathrm{S}$, which remains unchanged with the pose, the network can focus on modeling facial expressions.
Meanwhile, as $\mathrm{S}$ is disentangled into mean face template $\mathrm{\overline{S}}$ and the deformation $\mathrm{D}$, only the zero-centered $\mathrm{D}$ is required to be predicted, thereby reducing the complexity of the fitting process.

To facilitate the self-alignment process, the face geometry $\mathrm{G}$, pose-invariant face $\mathrm{S}$, mean face $\mathrm{\overline{S}}$, deformation $\mathrm{D}$, and pose-dependent face $\mathrm{P}$ are transformed into UV space~\cite{feng2018joint} as UV maps.
Since pixels with the same coordinates in the UV maps correspond to the same semantic region, the self-alignment module estimates the similarity transformation matrices using two sets of landmarks extracted from $\mathrm{P}$ and $\mathrm{S}$, respectively, based on the same UV coordinate set.

The dual face regression network is pretrained in a supervised manner on the 300W-LP dataset~\cite{zhu2016face}.
After the pretraining, we select the branch of the network that generates the pose-invariant face and employ it as our pose normalizer. This component takes a facial image as input and generates the corresponding pose-invariant face.

\section{Experimental Setup}
\subsection{Dataset}
The experiments are carried out on LRS3~\cite{afouras2018lrs3} and MEAD~\cite{wang2020mead} dataset. 
LRS3 is a large-scale in-the-wild dataset that includes videos obtained from the TED YouTube channel. MEAD contains talking-face videos which are simultaneously recorded at seven different views in a strictly-controlled environment.

In our experiments, we simulate two-speaker mixtures from these two datasets, respectively. The target speech is mixed with a random interference speech at a random signal-to-noise ratio (SNR) in the range from -10 dB to 10 dB. The audio sampling rate is 16 kHz and the face track video of the target speaker is provided at 25 frames-per-second (FPS).
For LRS3-2mix, there are 20,000 and 5,000 speech mixture utterances for training and validation simulated from the trainval set, and 3,000 speech mixtures for testing from 168 unseen and unheard speakers during training.
For MEAD-2mix, we only select videos with neutral emotion intensity captured from the frontal view for training and validation. The test set contains videos from seven different views, as shown in Figure~\ref{fig:overview}-(c).
In total, it consists of 10,000, 1,000 speech mixture utterances for training and validation, and 1,000 for each individual view for testing.

\subsection{Implementation details}
On LRS3-2mix, we train the whole neural network for 50 epochs with the initial learning rate set to $1e^{-3}$.
On MEAD-2mix, we load the weights of the pretrained model on LRS3 to get a good starting point and fine-tune the neural network for 20 epochs with the initial learning rate set to $1e^{-4}$.
The learning rate is halved if the accuracy on the validation set does not improve for 3 consecutive epochs. The training process would stop if the accuracy does not improve for 6 consecutive epochs. Adam~\cite{kingma2015adam} is used as the optimizer. Gradient clipping with a maximum L2-norm of 5 is applied during training.

\section{Results}

\begin{table*}[t]
\centering
\caption{Pose-invariant evaluation on MEAD dataset. Performance is reported with SDR(dB). Different views correspond to Figure \ref{fig:overview}-(c). PF refers to input with pose-invariant faces, and the masking regions in PIAVE (Mask Lip) and PIAVE (Mask Upper) is illustrated in Figure \ref{fig:overview}-(b). Avg(7) refers to the average SDR value of all 7 different views. Avg(6) refers to the average SDR value of 6 different views except for the front view.}
\begin{tabular}{lccccccccc}
\hline
\textbf{Model} & \textbf{Front} & \textbf{Top} & \textbf{Down} & \textbf{Left\_30} & \textbf{Left\_60} & \textbf{Right\_30} & \textbf{Right\_60} & \textbf{Avg(7)} & \textbf{Avg(6)}\\ \hline
PIAVE~(w/o PF) & 9.712  & 5.641 & 5.698 & 4.888 & 1.875 & 7.226 & 4.532  & 5.653 & 4.977 \\
PIAVE~(Mask Lip) & 10.277 & 7.078 & 5.301 & 5.328 & 5.804 & 5.277 & 5.107 & 6.310 & 5.649 \\
PIAVE~(Mask Upper) & 9.974 & 6.615 & 6.718 & 8.102 & 5.951 & 8.170 & \textbf{6.142} & 7.382 & 6.950  \\
PIAVE & \textbf{11.773} & \textbf{8.923} & \textbf{8.514} & \textbf{8.583}    & \textbf{6.118} & \textbf{8.387}  & 4.935 & \textbf{8.176} & \textbf{7.577} \\ \hline
\end{tabular}
\label{tb:multi_view}
\end{table*}

\begin{table}[htbp]
\centering
\caption{Ablation study on LRS3 dataset. PF refers to input with pose-invariant faces, and the masking regions in PIAVE (Mask Lip) and PIAVE (Mask Upper) is illustrated in Figure \ref{fig:overview}-(b).}
\begin{tabular}{lccccc}
\hline
\textbf{Sys.} & \textbf{SI-SDR} & \textbf{SDR}    & \textbf{PESQ}  & \textbf{STOI}\\ \hline
PIAVE~(w/o PF) & 14.734          & 15.064          & 2.442 & 0.937 \\
PIAVE~(Mask Lip) & 14.573          & 14.906          & 2.461 & 0.937 \\
PIAVE~(Mask Upper) & 14.970  & 15.279 & 2.544 & 0.942 \\
PIAVE & \textbf{15.255} & \textbf{15.569} & \textbf{2.585} & \textbf{0.944} \\ \hline
\end{tabular}
\label{tb:ablation}
\end{table}

\begin{table}[htbp]
\centering
\caption{Comparison of our PIAVE model with the baseline models on LRS3 dataset. }
\begin{tabular}{lcccc}
\hline
\textbf{Model} &  \textbf{SI-SDR} & \textbf{SDR} & \textbf{PESQ} & \textbf{STOI}\\ \hline
Visualvoice~\cite{gao2021visualvoice} & 9.603 & 10.112 & 2.089 & 0.897 \\
TDSE~\cite{wu2019time} & 13.253 & 13.778 & 2.351 & 0.912 \\
PIAVE (Ours) & 15.255 & 15.569 & 2.585 & 0.944 \\ \hline
\end{tabular}
\label{tb:baseline}
\end{table}

\subsection{Pose-invariant evaluation on MEAD}
To analyze the effectiveness of pose-invariant faces in AVSE, we evaluate several variants of PIAVE as shown in Table\ref{tb:multi_view}.
One major impact of pose variations on visually-assisted speech processing systems is decreased performance in mismatched train/test conditions, i.e., the neural network is trained and tested on different poses~\cite{petridis2017end}.
Therefore, we conduct evaluations on the MEAD dataset, which includes videos captured from multiple views simultaneously, as illustrated in Figure \ref{fig:overview}-(c), to assess the performance of different systems under mismatched conditions.
Specifically, we train the system using only front-view videos and test it on seven different views.

In Table~\ref{tb:multi_view}, we use the signal-to-distortion ratio (SDR) to measure the signal quality of the extracted speech under multiple views.
For pose-invariant evaluation, we used the average SDR across all seven views (Avg(7)) to measure the overall signal quality and Avg(6) to measure the signal quality under mismatched train/test conditions.
Our results indicate that the performance of the model degrades as the view distance from the front one increases, demonstrating the impact of pose variations.
PIAVE (w/o PF) performs the worst among all variants, indicating the severe degradation of performance without the pose-invariant faces under pose variations.
Comparison between PIAVE (Mask Upper) and PIAVE (Mask Lip) reveals that the movement of lip region have a more significant impact than the upper region of pose-invariant faces.
The integration of both components, i.e., the entire face region, yields the best performance in PIAVE, with a $45\%$ average improvement in performance over PIAVE (w/o PF).
In conclusion, introducing pose-invariant faces improves the system's performance and makes it more robust against mismatched conditions under pose variations, bringing it one step closer to pose-invariant AVSE.

\subsection{Ablation study of PIAVE on LRS3}
\label{subsection:ablation}
In Table~\ref{tb:ablation}, we present an ablation study on LRS3, a large-scale in-the-wild dataset, to elaborate on the significance of different video streams.
Besides SI-SDR and SDR to measure the signal quality, we use the perceptual evaluation of speech quality (PESQ)~\cite{rix2001perceptual} and the short term objective intelligibility (STOI)~\cite{taal2010short} to evaluate the perceptual quality and intelligibility of the extracted speech. The higher the better for these metrics.

To illustrate the effect of incorporating pose-invariant faces, PIAVE brings about 0.52dB SI-SDR and 0.143 PESQ improvement when compared to PIAVE (w/o PF).
When the lip region of pose-invariant faces is masked in PIAVE (Mask Lip), the resulting performance is even worse than that of PIAVE (w/o PF) in terms of SI-SDR and SDR. This result highlights that primarily the lip movements of pose-invariant faces contribute to the performance improvement.
Additionally, PIAVE (Mask Upper) outperforms PIAVE (w/o PF), but there is still a gap between PIAVE (Mask Upper) and PIAVE, emphasizing the need for the utilization of the entire face region.
Specifically, the upper region of the talking face offers a coarse-grained synchronization cue, while the lip region provides fine-grained viseme-phoneme mapping. We note that when the amount of training data is limited, the benefits of the upper region of the talking face can be larger, as depicted in Table~\ref{tb:multi_view}.
In summary, our results show that for the dataset with various pose variations, providing the model with stable pose-invariant visual input helps machine listening. Furthermore, integrating pose-invariant faces and original face tracks results in an improved observation of the talking face from multiple views, including lip movements and facial expressions from the upper region of the face, in which lip movements play a major role.

\subsection{PIAVE vs baseline on LRS3}
In Table~\ref{tb:baseline}, a comparison of the results of our PIAVE model with two recent AVSE models, VisualVoice~\cite{gao2021visualvoice} and TDSE~\cite{wu2019time}, is presented.
VisualVoice is a complex spectral mapping approach that is used for speech separation, and TDSE is a time-domain speaker extraction system.
We reproduce the aforementioned two baseline models and evaluated them on the same dataset used for training and testing PIAVE, to ensure a fair comparison. The results demonstrate that PIAVE outperforms the current SOTA models in terms of speech quality and intelligibility.

\section{Discussion}
Head pose variations affect the identification of the auditory component of audio-visual speech stimuli~\cite{jordan1997seeing}, and how to reduce their impact has been the focus of previous research.
This includes attempts to develop pose-invariant lipreading through the learning of multiple classifiers for each specific poses~\cite{hesse2012multi} or with a mapping function to transform features to a specific view, e.g. $30^{\circ}$, $45^{\circ}$ and $60^{\circ}$ to the frontal view~\cite{estellers2011multipose} or to the $30^{\circ}$ view~\cite{lan2012view}.
However, these approaches have their limitations in terms of being time-consuming with the increase of pose numbers and constrained by specific pose orientations.
Recent research in lipreading has explored the efficacy of using multiple views of lips for improved performance~\cite{petridis2017end}. While this approach has shown to be beneficial, it is limited by the need for multiple cameras.
Moreover, recent research has investigated the impact of head movements on audio-visual speech enhancement and demonstrates the effectiveness of using face frontalization to remove head movements~\cite{kang2022impact}. However, this approach is only validated on frontal views.

We present the first attempt to address the pose variation problem in audio-visual speaker extraction with a pose-invariant view. Unlike previous approaches to this problem, we generate a corresponding pose-invariant view for any visual input with only one camera to enable stable visual input and multi-view observation. We believe this work can provide inspiration for other visually-assisted speech processing algorithms.

There are also many open future directions. In the current stage, the pose-invariant face generated by the pose normalizer lacks facial texture, which typically contains identity information such as gender. This limitation may offset the advantages of pose normalization as a cue for the model to accurately identify the target speaker.
Furthermore, potential areas for future research also include exploring more effective techniques for feature fusion between two video streams, as well as between audio and visual modality.

\section{Conclusion}
In this paper, we have proposed a Pose-invariant Audio-visual Speaker Extraction Network (PIAVE) to address the pose variation problem, which is largely unexplored in AVSE.
Specifically, we generate the pose-invariant view from each original pose orientation,
which enables the model to receive a consistent frontal view of the talker regardless of his/her head pose.
We validate the effectiveness of PIAVE on two datasets: the large-scale in-the-wild LRS3 dataset and the multi-view talking face MEAD dataset.
Our experimental results demonstrate that the incorporation of pose-invariant faces results in a more robust model capable of handling variations in the head pose. It also improves the overall performance, by enabling stable input of the visual modality, as well as multi-view observation of talking faces.
In summary, PIAVE provides a practical solution for the pose variation problem and is a step forward in modeling the cocktail party effect in uncontrollable circumstances.

\section{Acknowledgments}
\label{sec:Acknowledgments}
This work is supported by
1) Huawei Noah’s Ark Lab;
2) National Natural Science Foundation of China (Grant No. 62271432);
3) Guangdong Provincial Key Laboratory of Big Data Computing, The Chinese University of Hong Kong, Shenzhen (Grant No. B10120210117-KP02);
4) German Research Foundation (DFG) under Germany's Excellence Strategy (University Allowance, EXC 2077, University of Bremen).

\bibliographystyle{IEEEtran}
\bibliography{mybib}

\end{document}